\documentclass[aps,physrev,reprint,twocolumn,superscriptaddress,10pt,amsmath,showkeys]{revtex4-2}
\usepackage[T1]{fontenc}
\usepackage{mathtools,newtxtext}

\mathtoolsset{showonlyrefs}
\usepackage{newtxmath}
\usepackage{graphicx,booktabs,xcolor}

\usepackage[breaklinks]{hyperref}
\hypersetup{pdftitle={Temporal Portability of Numeric User Metadata on Twitter},colorlinks,citecolor=[rgb]{.094,.223,.617},urlcolor=[rgb]{.094,.223,.617},linkcolor=[rgb]{.094,.223,.617}}

\begin{document}

\title{Temporal Portability of Numeric User Metadata on Twitter}
\author{Chako Takahashi}
\email{c.takahashi@scu.ac.jp}
\affiliation{Advanced Intelligence Technology Center, Sapporo City University, 9th Floor, Jobkita Building, Minami 1-jo Nishi 6-chome 20-1, Chuo-ku, Sapporo, Hokkaido 060-0061, Japan}
\author{Mitsuo Yoshida}
\affiliation{Institute of Business Sciences, University of Tsukuba, 3-29-1 Otsuka, Bunkyo-ku, Tokyo 112-0012, Japan}
\author{Muneki Yasuda}
\affiliation{Graduate School of Science and Engineering, Yamagata University, 4-3-16 Jonan, Yonezawa, Yamagata 992-8510, Japan}

\begin{abstract}
Numeric user metadata in social media are often reused over time.
However, their reusability may depend on what an analysis needs to preserve.
We introduce \emph{temporal portability} as an analytical perspective for assessing the cross-time reuse of user features and feature-based rules.
Specifically, we ask how well relevant properties are preserved when features and rules defined at a source time point are reused at a target time point.
We used quarterly data on user features obtained directly from or derived from Japanese-language tweets in Twitter's 1\% sample stream from 2020-Q1 to 2022-Q3.
Each quarter included approximately 10.1--11.0 million unique users.
We evaluated 13 numeric user features in terms of feature distributions, same-user relative ranks, selection rates, and selected-user membership.
Across quarters, feature distributions changed and, for many features, same-user relative ranks were less well preserved at longer quarter lags.
Reusing source-quarter thresholds also produced selection-rate drift.
Target-quarter recalibration nearly matched source-quarter selection rates.
However, membership turnover persisted and increased at longer quarter lags.
Our results show that temporal portability should be assessed in terms of the property that an analysis needs to preserve.
\end{abstract}


\maketitle

\section{Introduction}
\label{sec:introduction}

On Twitter and other social media, numeric user metadata associated with users or accounts, such as numbers of followers, friends, posts, and favourites, are used in many user-level analyses.
For example, these metadata are used to assess user influence \cite{Cha2010Influence}, infer demographic characteristics \cite{Sloan2015WhoTweets}, estimate location \cite{Hironaka2021CrossCountry}, and detect bots \cite{Yang2020BotDetection}.
Compared with reconstructing large-scale networks or extracting features from large volumes of post content, obtaining metadata directly from profile or account objects can reduce the burden of data collection and processing \cite{Cresci2015FameForSale,Yang2020BotDetection}.
Similar account-level metadata have also been used on other social media, including Instagram \cite{Tricomi2023Crowdturfing}, Reddit \cite{Saeed2022TrollMagnifier}, and YouTube \cite{RibeiroWest2021YouNiverse}.
Analyses that compare user populations across periods may apply a user feature from an earlier period, or a feature-based threshold or selection rule calibrated using earlier data, to a later period so that the populations are compared using the same criterion.
Such cross-time reuse does not necessarily preserve the property of interest from one period to another.

Cross-time reuse may be affected both by changes among the same users and by changes in the composition of the user population observed in each period.
The activity patterns of individual users change over time \cite{Liu2014TweetsChanging}, and their profile information is updated \cite{Shima2017ProfileChanges}.
The temporal pattern of a user activity may also depend on how it is represented, for example as a raw count or a normalised rate \cite{Garimella2021RetweetCareers}.
The composition of the observed user population likewise changes across periods.
On Twitter, successive user cohorts have been reported to differ in activity and interaction patterns \cite{Wolf2022SuccessiveCohorts}.
On Reddit, the influx of new and newly political users has been shown to contribute substantially to changes in platform-level political polarisation \cite{WallerAnderson2021Polarization}.
Both types of change may affect the results obtained when the same metadata or rule is used in different periods.

Research on cross-time reuse in social media has examined how model performance, user ranking, selected-user membership, and threshold-based classification are preserved over time in specific contexts.
Bot-detection studies have assessed the effectiveness of previously used features for new bot populations \cite{DeNicola2021OldFeatures} and classifier generalisation across different bot populations \cite{Echeverria2018LOBO}.
Research on natural language processing for social media has also examined adaptation to temporal distribution shift \cite{Mireshghallah2023TemporalAdaptation} and temporal generalisation \cite{Ushio2024TemporalGeneralization}.
Studies of user ranking have analysed temporal influence rankings \cite{Ma2017TemporalInfluence} and reported changes in relative ranking and top-percentile membership across political events \cite{LiangLu2023Influencers}.
Users identified as influential in one period do not necessarily remain influential in later periods \cite{Yamada2025StableInfluencers}, and changes in scores over time can alter classification results obtained using a fixed numeric threshold \cite{RauchfleischKaiser2020FalsePositive}.
However, these studies typically examine temporal change in relation to a particular analytical objective.
It remains unclear how the assessment of cross-time reusability changes according to the property that an analysis needs to preserve.
A single measure of temporal change may provide an incomplete basis for deciding whether a feature or rule can be reused in a later period.

We introduce \emph{temporal portability} as an analytical perspective for assessing cross-time reuse.
We define temporal portability as the extent to which a property that an analysis aims to preserve is maintained when a user feature or feature-based user-selection rule used in a source period is reused in a target period.
This perspective concerns the cross-time reuse of features and rules within the broader question of temporal validity, which refers to whether knowledge and findings remain valid across time \cite{Munger2023TemporalValidity}.
Specifically, we assess temporal portability with respect to four properties: feature distributions, same-user relative ranks, selection rates, and selected-user membership.
Feature distributions refer to the distributions observed in the user population in each period, whereas same-user relative ranks refer to the relative ordering of users observed in both periods.
Selection rates refer to the proportion of users selected when a numeric threshold defined in the source period is applied in the target period.
Selected-user membership refers to which users are selected after the threshold is recalibrated in the target period so that the selection rate nearly matches that in the source period.
These four properties capture different aspects of temporal portability.
For example, preserving the selection rate does not necessarily preserve selected-user membership.
High temporal portability from one property therefore does not imply that another is preserved to the same extent.

We assess the temporal portability of 13 user features obtained from Twitter user metadata and the user-selection rules based on them.
By evaluating the same features and the associated selection rules across the four properties defined above, we provide a systematic comparison of how temporal portability depends on what an analysis seeks to preserve.
Our data comprise numeric feature values for approximately 10.1--11.0 million unique users observed per quarter through Japanese-language tweets collected from Twitter's 1\% sample stream over the 11 quarters from 2020-Q1 to 2022-Q3.
We first examine how feature distributions and same-user relative ranks differ between source and target quarters and how these differences vary with quarter lag.
We then examine how much the selection rate differs when an absolute numeric threshold defined in the source quarter is reused in the target quarter.
Finally, we compare how well selected-user membership is preserved when the target-quarter threshold is recalibrated so that the selection rate nearly matches that in the source quarter.

We found that temporal portability varied across the properties assessed.
In particular, even when target-quarter recalibration made the selection rate nearly match that in the source quarter, source-selected turnover persisted for many features.
This result indicates that matching the selection rate does not necessarily mean selecting the same users.
Which features showed higher temporal portability also depended on the property assessed.
Our results show that temporal portability should be assessed in terms of the property that a subsequent analysis needs to preserve.
By characterising these differences across 13 user features, we provide an empirical reference for researchers considering whether and how to reuse user metadata and selection rules across periods.

\section{Data}
\label{sec:data}

\subsection{Data collection}
\label{sec:data-collection}

We analysed tweets labelled as Japanese (\texttt{language=ja}) that were collected from the 1\% sample stream of the Twitter Streaming API (\texttt{statuses/sample}).
The observation period comprised 33 months, from 1 January 2020 to 30 September 2022.
We divided this period into calendar quarters: Q1 (January--March), Q2 (April--June), Q3 (July--September), and Q4 (October--December).
For each quarter, we constructed the observed user set from the user objects attached to tweets observed in that quarter.
When multiple tweets from the same user were observed within a quarter, we used the user object attached to the last observed tweet for that user.

We collected 708,331,124 tweets over the 33-month observation period.
Each quarter contained approximately 61--72 million tweets from approximately 10.1--11.0 million unique users.
Table~\ref{tab:quarterly-data} in Appendix~\ref{app:data-user-set-sizes} presents the detailed quarterly counts of tweets and unique users.

\subsection{User features}
\label{sec:user-features}

\begin{table*}[tb]
\caption{User features.}
\label{tab:user-features}
\centering\footnotesize
\resizebox{\textwidth}{!}{%
\begin{tabular}{@{\hspace{8pt}}llll@{\hspace{8pt}}}
\toprule
Feature name & Description & Calculation & Type \\
\midrule
\textbf{Count features} & & & \\
\texttt{user.friends\_count} & Number of accounts followed (friends) &  & count \\
\texttt{user.followers\_count} & Number of followers &  & count \\
\texttt{user.listed\_count} & Number of lists containing the user &  & count \\
\texttt{user.statuses\_count} & Number of tweets &  & count \\
\texttt{user.favourites\_count} & Number of favourite tweets &  & count \\
\midrule
\textbf{Derived features} & & & \\
\texttt{account\_age\_days} & Days since account creation & $\max\bigl(\text{days}(\texttt{created\_at} - \texttt{user.created\_at}),\, 1\bigr)$ & age \\
\texttt{friends\_rate} & Number of accounts followed per day of account age & $\texttt{user.friends\_count} / \texttt{account\_age\_days}$ & rate \\
\texttt{followers\_rate} & Number of followers per day of account age & $\texttt{user.followers\_count} / \texttt{account\_age\_days}$ & rate \\
\texttt{listed\_rate} & Number of lists containing the user per day of account age & $\texttt{user.listed\_count} / \texttt{account\_age\_days}$ & rate \\
\texttt{tweet\_rate} & Number of tweets per day of account age & $\texttt{user.statuses\_count} / \texttt{account\_age\_days}$ & rate \\
\texttt{favourites\_rate} & Number of favourite tweets per day of account age & $\texttt{user.favourites\_count} / \texttt{account\_age\_days}$ & rate \\
\texttt{followers\_per\_tweet} & Follower count relative to tweet count & $\texttt{user.followers\_count} / \max(\texttt{user.statuses\_count}, 1)$ & ratio \\
\texttt{followers\_friends\_ratio} & Followers-to-friends ratio & $(\texttt{user.followers\_count} + 1) / (\texttt{user.friends\_count} + 1)$ & ratio \\
\bottomrule
\end{tabular}%
}
\end{table*}

For each user, we used five count features obtained directly from the Twitter user object and eight features derived from them.
We refer to these 13 features collectively as \emph{user features}.
Table~\ref{tab:user-features} lists the features and their definitions.
The five numeric user metadata fields obtained directly from the user object were the number of accounts followed (\texttt{user.friends\_count}), the number of followers (\texttt{user.followers\_count}), the number of lists containing the user (\texttt{user.listed\_count}), the number of tweets (\texttt{user.statuses\_count}), and the number of favourite tweets (\texttt{user.favourites\_count}).
Using the account creation time \texttt{user.created\_at} in the user object and the posting time \texttt{created\_at} recorded in the tweet object, we define the account age in days at observation, \texttt{account\_age\_days}, as $\texttt{account\_age\_days} = \max\bigl(\text{days}(\texttt{created\_at} - \texttt{user.created\_at}),\, 1\bigr)$.
To represent user activity while accounting for the duration of account use, we used five account-age-normalised rates obtained by dividing the count features by \texttt{account\_age\_days}: \texttt{friends\_rate}, \texttt{followers\_rate}, \texttt{listed\_rate}, \texttt{tweet\_rate}, and \texttt{favourites\_rate}.
We also used \texttt{followers\_per\_tweet}, which represents follower count relative to tweet count, and \texttt{followers\_friends\_ratio}, which represents follower count relative to the number of accounts followed.
Table~\ref{tab:user-features} provides descriptions and definitions for all features.
Let \(\mathcal{F}\) denote the set of these 13 user features.

\subsection{Temporal comparisons and user sets}
\label{sec:user-sets}

Let the chronologically ordered set of the 11 calendar quarters from 2020-Q1 to 2022-Q3 be
\begin{align}
\mathcal{T}=\{\text{2020-Q1},\text{2020-Q2},\ldots,\text{2022-Q3}\}.
\label{eq:quarter-set}
\end{align}
For each quarter $\tau\in\mathcal{T}$, let $U_\tau$ denote the set of unique users observed in that quarter.
Table~\ref{tab:quarterly-data} in Appendix~\ref{app:data-user-set-sizes} presents the cardinality \(|U_{\tau}|\) as observed users.

For comparisons between distinct quarters, we consider \(s,t\in\mathcal{T}\) with \(s<t\), where $s<t$ means that $s$ precedes $t$.
We refer to the earlier quarter $s$ as the source quarter and the later quarter $t$ as the target quarter.
We define the number of quarterly intervals from source quarter $s$ to target quarter $t$ as the quarter lag $\operatorname{lag}(s,t)$, denoted by $\ell\in\{1,\ldots,10\}$, with
\begin{align}
\operatorname{lag}(s,t)=\ell.
\label{eq:lag}
\end{align}
Let the set of source--target quarter pairs with quarter lag $\ell$ be
\begin{align}
\mathcal{P}_{\ell}=\left\{(s,t)\in\mathcal{T}^2\mid s<t,\ \operatorname{lag}(s,t)=\ell\right\}.
\label{eq:lag-pair-set}
\end{align}
For example, $\mathcal{P}_{1}$ contains the 10 adjacent quarter pairs.
The set $\mathcal{P}_{10}$ consists of the single pair with 2020-Q1 as the source quarter and 2022-Q3 as the target quarter.

We define the set of users observed in both quarters \(s\) and \(t\) as
\begin{align}
C_{s,t}=U_s\cap U_t.
\label{eq:common-user-set}
\end{align}
Across the 55 source--target quarter pairs formed from the 11 quarters, each pairwise-common user set \(C_{s,t}\) contains approximately 3.29--6.41 million users.
Table~\ref{tab:common-user-counts} in Appendix~\ref{app:data-user-set-sizes} presents the cardinalities $|C_{s,t}|$.

\section{Measures of Temporal Portability}
\label{sec:portability-measures}

We assess the temporal portability of user features and the user-selection rules based on them from four properties: feature distributions, same-user relative ranks, selection rates, and selected-user membership.
We use the $L_1$ distance for feature distributions, cross-time Spearman rank correlation for same-user relative ranks, absolute selection-rate drift for selection rates, and source-selected turnover for selected-user membership.
Analyses of all users observed in each quarter use \(U_\tau\) or \(U_s,U_t\), whereas analyses that match users between the source and target quarters use the pairwise-common user set \(C_{s,t}\).

\subsection{Feature distributions}
\label{sec:measure-l1}

We assess differences in the distribution of each user feature over time.
We represent the marginal distribution in each quarter as a histogram and quantify distributional differences by comparing the histograms for source quarter $s$ and target quarter $t$.

For each quarter, we randomly sampled $N$ users without replacement from the observed user set \(U_\tau\) and repeated the procedure $K$ times.
Let the histogram for feature \(i\in\mathcal{F}\), obtained from the sample drawn in repeat \(k\) for quarter \(\tau\), be
\begin{align}
\mathbf{h}^{(k)}_{\tau,i} = \left( h^{(k)}_{\tau,i,1},\ldots,h^{(k)}_{\tau,i,B_i} \right),
\label{eq:histogram-vector}
\end{align}
where \(B_i\) is the effective number of bins for feature \(i\).
For each feature, we used common bin boundaries across all quarters and normalised the bin values such that $\sum_{b=1}^{B_i}h_{\tau,i,b}^{(k)}=1$.

We define the difference between the histograms for quarters $s$ and $t$ as the $L_1$ (Manhattan) distance:
\begin{align}
d\left(\mathbf{h}^{(k)}_{s,i},\mathbf{h}^{(k')}_{t,i}\right) = \sum_{b=1}^{B_i} \left| h^{(k)}_{t,i,b} - h^{(k')}_{s,i,b} \right|.
\label{eq:l1-distance}
\end{align}
We define the between-quarter distance as the average distance between samples from quarters $s$ and $t$:
\begin{align}
D_i^{\mathrm{between}}(s,t) &= \frac{1}{K^2} \sum_{k=1}^{K}\sum_{k'=1}^{K} d\left(\mathbf{h}^{(k)}_{s,i},\mathbf{h}^{(k')}_{t,i}\right).
\label{eq:between-quarter-distance}
\end{align}

Because finite samples produce differences between histograms even within a quarter, we define the within-quarter distance to quantify the magnitude of this sampling variation:
\begin{align}
D_i^{\mathrm{within}}(\tau) &= \frac{1}{K(K-1)} \sum_{k\neq k'} d\left(\mathbf{h}^{(k)}_{\tau,i},\mathbf{h}^{(k')}_{\tau,i}\right).
\label{eq:within-quarter-distance}
\end{align}

In Section~\ref{sec:results-marginal-distributions}, we used a sample size of $N=500{,}000$ users and $K=10$ repeats.
We applied the log transformation \(\log(1+x)\) to each feature before constructing the histograms. For each feature, we determined quantile-based boundaries for 100 bins from the transformed values across all 11 quarters and used the same boundaries for every quarter.
Because we removed duplicate boundaries, \(B_i\) can be less than 100 for some features. For each quarter lag \(\ell\), we aggregated the between-quarter distances in Equation~\eqref{eq:between-quarter-distance} over \((s,t)\in\mathcal{P}_{\ell}\), separately for each feature, and assessed how distributional differences varied with temporal separation.
For each feature $i$, we also aggregated the within-quarter distances in Equation~\eqref{eq:within-quarter-distance} over the 11 quarters $\tau \in \mathcal{T}$ and used their median as a reference for the between-quarter distance.

\subsection{Cross-time rank correlation}
\label{sec:measure-rank}

Even when a feature distribution changes over time, the relative ordering of the same users may be preserved.
We therefore use cross-time rank correlation to assess how well users observed in both source quarter \(s\) and target quarter \(t\) preserve their relative ordering for each feature.

For each source--target quarter pair \((s,t)\), we randomly sampled $N$ users without replacement from the pairwise-common user set \(C_{s,t}\) and repeated the procedure $K$ times.
We compared the feature values of the same sampled users between source quarter $s$ and target quarter $t$.
Users with tied feature values were assigned the mean rank for the tied positions.
Let the values of feature $i$ in the sample drawn in repeat $k$ for quarter $\tau$ be \(\mathbf{x}^{(k)}_{\tau,i} = \left(x^{(k)}_{\tau,i,1},\ldots,x^{(k)}_{\tau,i,N}\right)\).
We define the cross-time rank correlation between quarters $s$ and $t$ as the Spearman rank correlation:
\begin{align}
\rho^{\mathrm{rank},(k)}_{s,t,i}=\operatorname{Spearman}\left(\mathbf{x}^{(k)}_{s,i},\mathbf{x}^{(k)}_{t,i}\right).
\label{eq:cross-time-rank}
\end{align}
Values of \(\rho^{\mathrm{rank},(k)}_{s,t,i}\) closer to 1 indicate stronger preservation of the relative ordering of the same users for feature \(i\) between the two quarters.

In Section~\ref{sec:results-rank-correlations}, we used a sample size of \(N=500{,}000\) users and $K=3$ repeats.
For each source--target quarter pair and feature, we calculated the median of the $K$ cross-time rank correlations.
For each quarter lag \(\ell\), we then aggregated these median values from Equation~\eqref{eq:cross-time-rank} across the quarter pairs \((s,t)\in\mathcal{P}_{\ell}\) and assessed how the preservation of relative ranks varied with temporal separation.

\subsection{Selection-rate drift and recalibration}
\label{sec:measure-selection-drift}

We consider a feature-based user-selection rule that sets a threshold for each feature \(i\in\mathcal{F}\) and selects users whose values are at least that threshold.
We use absolute selection-rate drift to assess how much the proportion of selected users differed between source quarter \(s\) and target quarter \(t\) when a threshold defined in the source quarter was applied in the target quarter.

Let \(q\in(0,1)\) be the quantile level used to define the threshold.
For feature \(i\), the rule sets the threshold at the $q$-th quantile and selects users whose feature values are at least that threshold.

We first consider applying a source-quarter threshold to the observed user set in each quarter.
For repeat \(k\) of each source--target quarter pair \((s,t)\), we randomly drew disjoint training and evaluation samples of sizes \(N_{\mathrm{train}}\) and \(N_{\mathrm{eval}}\), respectively, from \(U_s\).
We used the training sample to calculate the threshold.
We also randomly drew a target-quarter evaluation sample of size \(N_{\mathrm{eval}}\) from \(U_t\). For feature \(i\), we define the source-quarter threshold from the training sample as
\begin{align}
\theta^{U,(k)}_{s,i}(q) = Q_q\left(\mathbf{x}^{U,\mathrm{train},(k)}_{s,i}\right),
\label{eq:fixed-threshold}
\end{align}
where \(Q_q(\cdot)\) denotes the sample $q$-th quantile.
We define the selection rates obtained by applying this threshold to the source-quarter and target-quarter evaluation samples, respectively, as
\begin{align}
\pi^{U,(k)}_{s,i}(q) &= \frac{1}{N_{\mathrm{eval}}} \sum_{n=1}^{N_{\mathrm{eval}}} \mathbb{I}\left[ x^{U,\mathrm{eval},(k)}_{s,i,n} \geq \theta^{U,(k)}_{s,i}(q) \right],
\label{eq:source-selection-rate}\\
\pi^{U,\mathrm{fixed},(k)}_{t|s,i}(q) &= \frac{1}{N_{\mathrm{eval}}} \sum_{n=1}^{N_{\mathrm{eval}}} \mathbb{I}\left[ x^{U,\mathrm{eval},(k)}_{t,i,n} \geq \theta^{U,(k)}_{s,i}(q) \right],
\label{eq:target-selection-rate}
\end{align}
where \(\mathbb{I}\;[\cdot]\) is the indicator function, which equals 1 when the condition in brackets is satisfied and 0 otherwise.
We define selection-rate drift as the change that occurs when the source-quarter threshold is held fixed and applied in the target quarter, and assess its absolute value:
\begin{align}
\left|\Delta\pi^{U,\mathrm{fixed},(k)}_{s,t,i}(q)\right| = \left| \pi^{U,\mathrm{fixed},(k)}_{t|s,i}(q) - \pi^{U,(k)}_{s,i}(q) \right|.
\label{eq:fixed-drift}
\end{align}

We next matched the evaluated users by user identifier between the source and target quarters and assessed the selection rate after threshold recalibration.
From each \(C_{s,t}\), we randomly drew disjoint training and evaluation sets of sizes $N_{\mathrm{train}}$ and $N_{\mathrm{eval}}$, respectively, using the same user identifiers in the source and target quarters. We define the source- and target-quarter thresholds from the corresponding values in the training set as
\begin{align}
\theta^{C,(k)}_{s,i}(q) &= Q_q\left(\mathbf{x}^{C,\mathrm{train},(k)}_{s,i}\right),
\label{eq:matched-threshold-s}\\
\theta^{C,(k)}_{t,i}(q) &= Q_q\left(\mathbf{x}^{C,\mathrm{train},(k)}_{t,i}\right).
\label{eq:matched-threshold-t}
\end{align}
We define the selection rates obtained by applying the source-quarter threshold in Equation~\eqref{eq:matched-threshold-s} to the source-quarter and target-quarter evaluation sets, respectively, as
\begin{align}
\pi^{C,(k)}_{s,i}(q) &= \frac{1}{N_{\mathrm{eval}}} \sum_{n=1}^{N_{\mathrm{eval}}} \mathbb{I}\left[ x^{C,\mathrm{eval},(k)}_{s,i,n} \geq \theta^{C,(k)}_{s,i}(q) \right],
\label{eq:matched-source-selection-rate}\\
\pi^{C,\mathrm{fixed},(k)}_{t|s,i}(q) &= \frac{1}{N_{\mathrm{eval}}} \sum_{n=1}^{N_{\mathrm{eval}}} \mathbb{I}\left[ x^{C,\mathrm{eval},(k)}_{t,i,n} \geq \theta^{C,(k)}_{s,i}(q) \right].
\label{eq:matched-fixed-selection-rate}
\end{align}
We define the recalibrated selection rate obtained by applying the target-quarter threshold in Equation~\eqref{eq:matched-threshold-t} to the target-quarter evaluation set as
\begin{align}
\pi^{C,\mathrm{recal},(k)}_{t,i}(q) = \frac{1}{N_{\mathrm{eval}}} \sum_{n=1}^{N_{\mathrm{eval}}} \mathbb{I}\left[ x^{C,\mathrm{eval},(k)}_{t,i,n} \geq \theta^{C,(k)}_{t,i}(q)\right].
\label{eq:matched-recal-selection-rate}
\end{align}
The corresponding absolute selection-rate drifts are defined as
\begin{align}
\left|\Delta\pi^{C,\mathrm{fixed},(k)}_{s,t,i}(q)\right| &= \left| \pi^{C,\mathrm{fixed},(k)}_{t|s,i}(q) - \pi^{C,(k)}_{s,i}(q) \right|,
\label{eq:matched-fixed-drift}\\
\left|\Delta\pi^{C,\mathrm{recal},(k)}_{s,t,i}(q)\right| &= \left| \pi^{C,\mathrm{recal},(k)}_{t,i}(q) - \pi^{C,(k)}_{s,i}(q) \right|.
\label{eq:matched-recal-drift}
\end{align}

For recalibration, we re-estimated the threshold at the same quantile level $q$ from the target-quarter training set and applied it to the target-quarter evaluation set.
Using the same quantile level in both quarters allowed us to compare selected-user membership under a common selection criterion.

In Section~\ref{sec:matched-cohort}, we used the quantile levels \(\mathcal{Q}=\{0.75,0.90,0.95\}\), which correspond to selecting approximately the top 25\%, 10\%, and 5\% of users by feature value, respectively.
We set \(N_{\mathrm{train}}=N_{\mathrm{eval}}=500{,}000\) users.
We used $K=10$ repeats to assess Equation~\eqref{eq:fixed-drift} with the quarter-specific observed user sets \(U_s, U_t\), and $K=3$ repeats to assess Equations~\eqref{eq:matched-fixed-drift} and \eqref{eq:matched-recal-drift} with the pairwise-common user set \(C_{s,t}\).
For each source--target quarter pair, feature, and quantile level, we calculated the median across the corresponding repeats.

\subsection{Source-selected turnover}
\label{sec:measure-turnover}

Even when selection is based on the same quantile level \(q\) after target-quarter recalibration, users selected in the source quarter are not necessarily selected in the target quarter.
We therefore assess source-selected turnover, defined as the proportion of users selected in the source quarter who were not selected after target-quarter recalibration.

We used the evaluation set and thresholds from the analysis of the pairwise-common user set \(C_{s,t}\) in Section~\ref{sec:measure-selection-drift}.

Let the sets of users selected when calculating the selection rates in Equations~\eqref{eq:matched-source-selection-rate} and \eqref{eq:matched-recal-selection-rate}, respectively, be
\begin{align}
S^{C,(k)}_{s,i}(q) &= \left\{n \mid x^{C,\mathrm{eval},(k)}_{s,i,n} \geq \theta^{C,(k)}_{s,i}(q) \right\},
\label{eq:source-selected-set}\\
S^{C,\mathrm{recal},(k)}_{t,i}(q) &= \left\{ n \mid x^{C,\mathrm{eval},(k)}_{t,i,n} \geq \theta^{C,(k)}_{t,i}(q) \right\}.
\label{eq:recal-selected-set}
\end{align}
We define source-selected turnover as the proportion of users selected in the source quarter who were not selected after target-quarter recalibration:
\begin{align}
T^{(k)}_{s\to t,i}(q) = \frac{\; \left| S^{C,(k)}_{s,i}(q) \setminus S^{C,\mathrm{recal},(k)}_{t,i}(q) \right| \;}{ \left| S^{C,(k)}_{s,i}(q) \right| }.
\label{eq:source-selected-turnover}
\end{align}

In Section~\ref{sec:membership}, we set \(N_{\mathrm{train}}=N_{\mathrm{eval}}=500{,}000\) users, \(K=3\), and \(q=0.90\).
For each source--target quarter pair and feature, we calculated the median source-selected turnover across the $K$ repeats.
For each quarter lag \(\ell\), we aggregated these values over \((s,t)\in\mathcal{P}_{\ell}\) separately for each feature.

\section{Results}
\label{sec:results}

\subsection{Feature distributions}
\label{sec:results-marginal-distributions}

\begin{figure*}[htb]
\centering
\includegraphics[width=\textwidth]{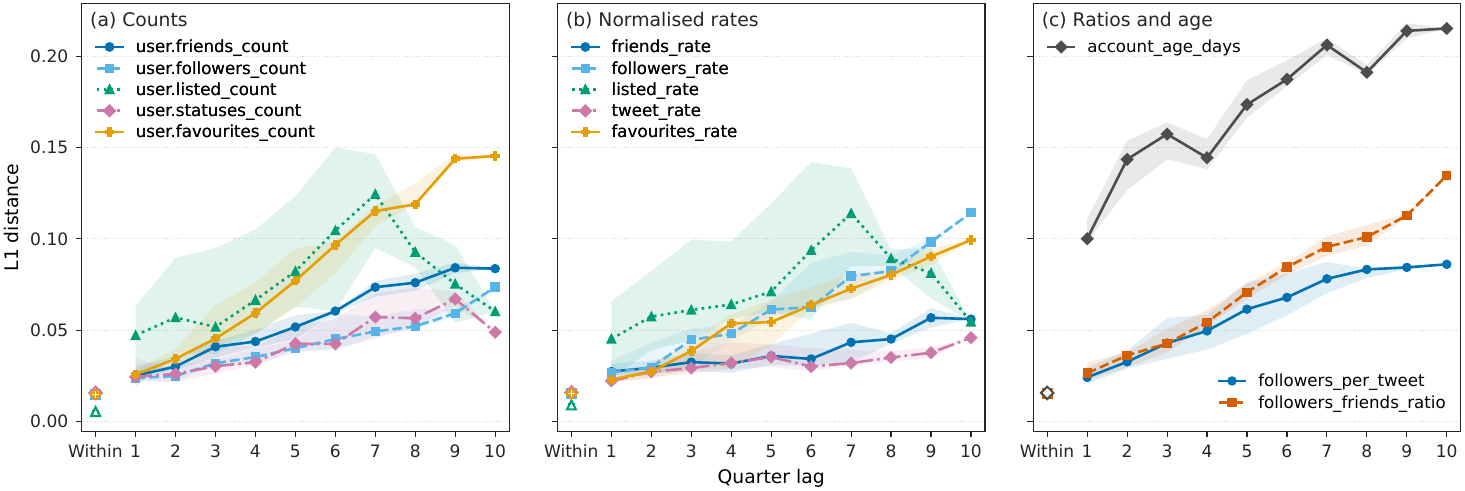}
\caption{$L_1$ distances between the marginal distributions of each user feature. Panel (a) shows the five count features in Table~\ref{tab:user-features}, (b) shows the five normalised rates obtained by normalising count features by account age (the features whose Type is \textit{rate} in Table~\ref{tab:user-features}), and (c) shows ratios and age. In each panel, Within, at the left of the horizontal axis, shows the median across the 11 quarters of the within-quarter distance \(D_i^{\mathrm{within}}(\tau)\) defined in Equation~\eqref{eq:within-quarter-distance}. For each quarter lag \(\ell\), points and lines show the median between-quarter distance \(D_i^{\mathrm{between}}(s,t)\) in Equation~\eqref{eq:between-quarter-distance} over \((s,t)\in\mathcal{P}_{\ell}\), and shaded regions show the interquartile range across quarter pairs. At \(\ell = 1\), the between-quarter distance exceeds the within-quarter reference for every feature, and many features have larger between-quarter distances at longer quarter lags. In contrast, \texttt{user.listed\_count} and \texttt{listed\_rate} peak at $\ell=7$, and \texttt{user.statuses\_count} peaks at $\ell=9$, producing non-monotonic patterns.}
\label{fig:marginal-l1}
\end{figure*}

Figure~\ref{fig:marginal-l1} presents the within-quarter distance for each user feature and the between-quarter distance by quarter lag $\ell$.
Figure~\ref{fig:marginal-l1}~(a) reports the five count features, Figure~\ref{fig:marginal-l1}~(b) the five normalised rates, and Figure~\ref{fig:marginal-l1}~(c) the two ratios and account age.
For all 13 features, the median between-quarter distance at \(\ell=1\), corresponding to adjacent quarters, exceeded the within-quarter reference.

For most features, the between-quarter distance increased almost monotonically with quarter lag, although the magnitude of this pattern differed by feature.
For \texttt{account\_age\_days} in Figure~\ref{fig:marginal-l1}~(c), the larger distances at longer quarter lags are clearly consistent with the feature's definition: account age increases as the observation period advances.
Using \texttt{account\_age\_days} as a reference case in which distributional differences are expected to be larger at longer lags, the increases in between-quarter distance from \(\ell=1\) to \(\ell=10\) were approximately 0.120 for \texttt{user.favourites\_count} and 0.110 for \texttt{followers\_friends\_ratio} in Figure~\ref{fig:marginal-l1}~(a).
These increases were comparable to the increase of 0.115 for \texttt{account\_age\_days}, suggesting that their marginal distributions varied over time to similar degrees.
By contrast, \texttt{friends\_rate} and \texttt{tweet\_rate} in Figure~\ref{fig:marginal-l1}~(b) showed more gradual changes in distance than the other features.

Three features showed non-monotonic relationships between distance and quarter lag.
The between-quarter distances for \texttt{user.listed\_count} in Figure~\ref{fig:marginal-l1}~(a) and \texttt{listed\_rate} in Figure~\ref{fig:marginal-l1}~(b) reached their maxima at $\ell=7$ and then declined.
The distance for \texttt{user.statuses\_count} in Figure~\ref{fig:marginal-l1}~(a) likewise reached its maximum at \(\ell=9\) before declining at \(\ell=10\).
The quarter-pair-level $L_1$ distances in Appendix~\ref{app:listed-features} are large for listed-related features in pairs containing 2021-Q4 or 2022-Q1.
For \texttt{user.statuses\_count}, relatively large distances occurred in pairs containing 2020-Q2.
Locally large distances for pairs containing a particular period can raise the aggregate for the corresponding lags, producing patterns such as those observed for these three features.

\subsection{Cross-time rank correlation}
\label{sec:results-rank-correlations}

\begin{figure*}[htb]
\centering
\includegraphics[width=\textwidth]{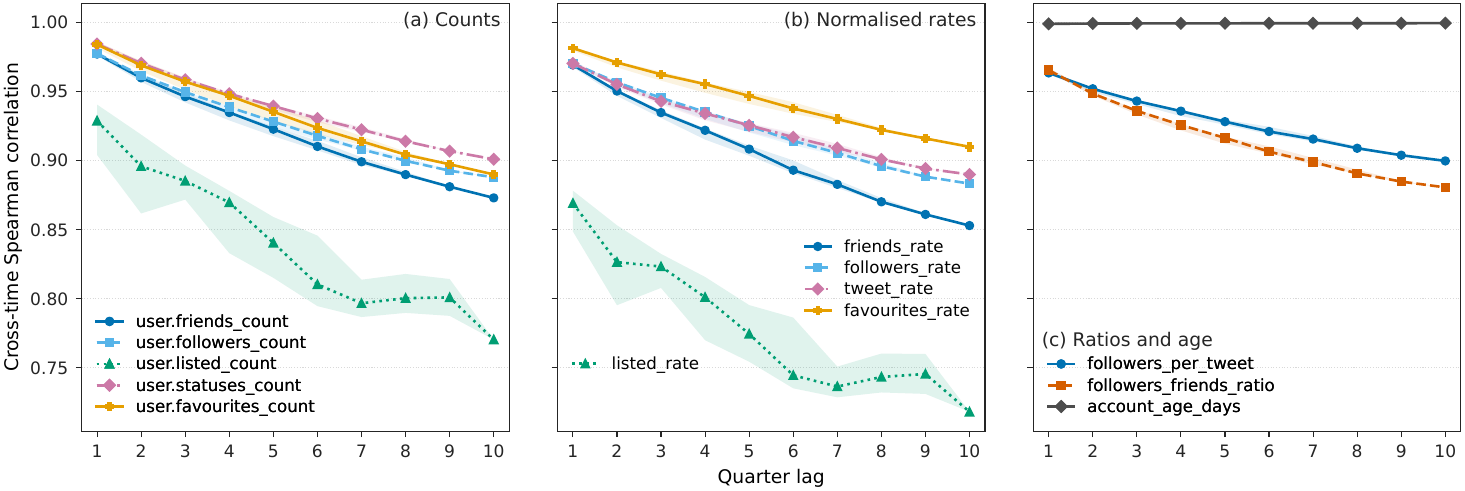}
\caption{Cross-time rank correlation among pairwise-common users \(C_{s,t}\).
Panel (a) shows the five count features in Table~\ref{tab:user-features}, (b) shows the five normalised rates, and (c) shows ratios and age.
For each quarter lag \(\ell\), points and lines show the median of Equation~\eqref{eq:cross-time-rank} obtained for each quarter pair over \((s,t)\in\mathcal{P}_{\ell}\), and shaded regions show the interquartile range across quarter pairs.
In panel (c), the rank correlation for \texttt{account\_age\_days} is close to 1 at every quarter lag, whereas the other 12 features tend to have lower rank correlations at longer lags.
The rank correlations for \texttt{user.listed\_count} in (a) and \texttt{listed\_rate} in (b) are lower than those for the other features and decline more at longer lags.}
\label{fig:cross-time-rank}
\end{figure*}

Figure~\ref{fig:cross-time-rank} presents the cross-time rank correlation for each user feature among pairwise-common users \(C_{s,t}\).
Figure~\ref{fig:cross-time-rank}~(a) reports the values of Equation~\eqref{eq:cross-time-rank} for the five count features, Figure~\ref{fig:cross-time-rank}~(b) those for the five normalised rates, and Figure~\ref{fig:cross-time-rank}~(c) those for the two ratios and account age.
All 12 features other than \texttt{account\_age\_days} tended to have lower cross-time rank correlations at \(\ell=10\) than at \(\ell=1\).
The feature \texttt{account\_age\_days} in Figure~\ref{fig:cross-time-rank}~(c) measures the elapsed time between account creation and observation for each user.
Its definition therefore implies that the relative ordering of the same users should remain largely preserved as the observation period advances.
The small variation in rank correlation across lags is consistent with this expected behaviour.
For the 10 features other than \texttt{user.listed\_count} and \texttt{listed\_rate} in Figures~\ref{fig:cross-time-rank}~(a) and (b), rank correlation declined gradually and almost monotonically from \(\ell=1\), reaching approximately 0.85--0.91 at \(\ell=10\).

The correlations for \texttt{user.listed\_count} in Figure~\ref{fig:cross-time-rank}~(a) and \texttt{listed\_rate} in Figure~\ref{fig:cross-time-rank}~(b) were already lower at $\ell=1$ than those for the other features and declined more at longer lags.
Many users had zero values for these two listed-related features, and the proportion of zero-valued users differed between quarters even when the analysis was restricted to the same users.
In each quarter, approximately 41--50\% of users had a value of zero for \texttt{user.listed\_count}, placing many users at the same rank.
When feature values are concentrated on a small number of integers, changes in \texttt{user.listed\_count} can readily move users into or out of a large tied group or alter their positions around it.
Such movements may be associated with the lower cross-time rank correlation.
Because Spearman rank correlation assigns the same rank to users with the same value, changes in the composition of these large tied groups may likewise be associated with the low cross-time rank correlations of the listed-related features.
Moreover, despite having zero values for the same users as \texttt{user.listed\_count}, \texttt{listed\_rate} had a lower rank correlation at every quarter lag.
Thus, alongside the many zero-valued users, changes in positive values and normalisation by account age may be associated with the low rank correlations of the listed-related features.

\subsection{Selection-rate drift and recalibration}
\label{sec:matched-cohort}

\begin{figure*}[htb]
\centering
\includegraphics[width=\textwidth]{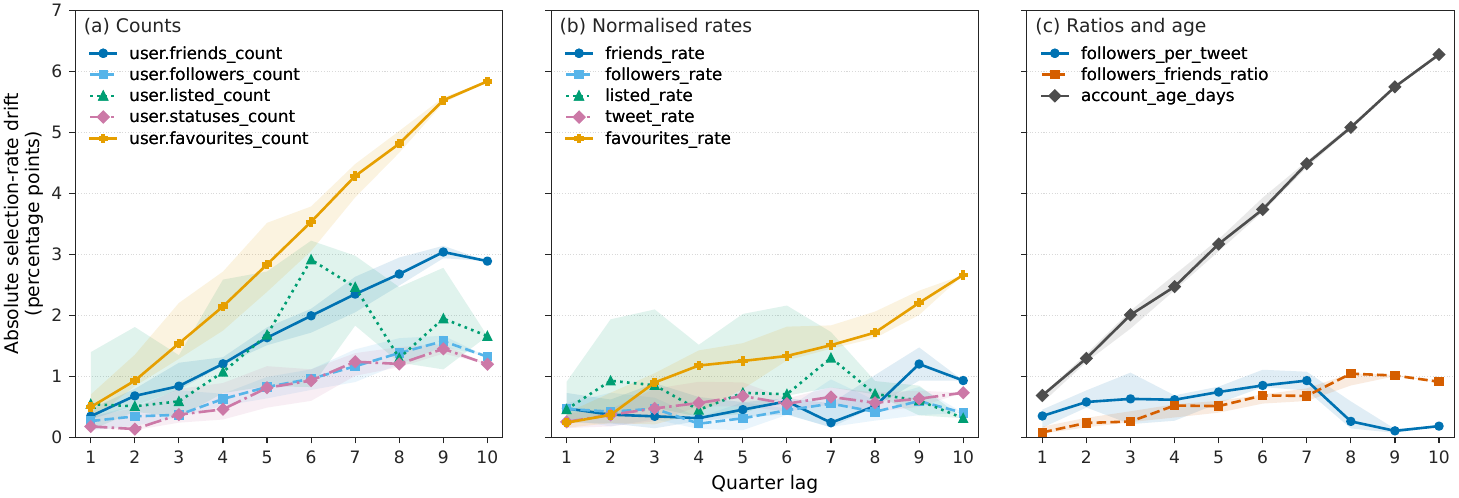}
\caption{Absolute selection-rate drift in Equation~\eqref{eq:fixed-drift} at \(q=0.90\) when the source-quarter threshold is held fixed and applied to the quarter-specific observed user sets \(U_s,U_t\).
Panel (a) shows the five count features in Table~\ref{tab:user-features}, (b) shows the five normalised rates, and (c) shows ratios and age.
For each quarter lag \(\ell\), points and lines show the median across quarter pairs, and shaded regions show the interquartile range.
Absolute selection-rate drift becomes markedly larger at longer quarter lags for \texttt{account\_age\_days} in (c) and \texttt{user.favourites\_count} in (a), with clear increases also observed for \texttt{user.friends\_count} and \texttt{favourites\_rate}.
Comparisons of features representing the same underlying user activity in (a) and (b) indicate that account-age-normalised rates tend to preserve the selection rate under a source-quarter threshold better than count features at longer quarter lags.}
\label{fig:selection-rate-portability}
\end{figure*}

Figure~\ref{fig:selection-rate-portability} presents the absolute selection-rate drift in Equation~\eqref{eq:fixed-drift} at \(q=0.90\) when the threshold defined in the source quarter was held fixed and applied to the quarter-specific observed user sets \(U_s,U_t\).
Figure~\ref{fig:selection-rate-portability}~(a) reports the five count features, Figure~\ref{fig:selection-rate-portability}~(b) the five normalised rates, and Figure~\ref{fig:selection-rate-portability}~(c) the two ratios and account age.
For many features, absolute selection-rate drift was larger at longer lags, but its magnitude and pattern differed clearly among features.

For \texttt{account\_age\_days} in Figure~\ref{fig:selection-rate-portability}~(c), absolute selection-rate drift increased from approximately 0.69 percentage points at \(\ell=1\) to approximately 6.27 percentage points at \(\ell=10\).
Treating \texttt{account\_age\_days}, whose value increases as the observation period advances, as a reference case in which the portability of a source-quarter threshold is expected to decline with lag, \texttt{user.favourites\_count} in Figure~\ref{fig:selection-rate-portability}~(a) showed a similarly large change, from approximately 0.50 percentage points at \(\ell=1\) to approximately 5.83 percentage points at \(\ell=10\).
The next largest increases were observed for \texttt{user.friends\_count} in Figure~\ref{fig:selection-rate-portability}~(a), whose absolute selection-rate drift increased from approximately 0.35 percentage points at \(\ell=1\) to approximately 2.89 percentage points at \(\ell=10\), and for \texttt{favourites\_rate} in Figure~\ref{fig:selection-rate-portability}~(b), for which it increased from approximately 0.24 percentage points to approximately 2.66 percentage points.
These increases were smaller than those for \texttt{account\_age\_days} and \texttt{user.favourites\_count}, but remained clear.

For the normalised rates in Figure~\ref{fig:selection-rate-portability}~(b) and the ratios in Figure~\ref{fig:selection-rate-portability}~(c), drift was generally smaller and was approximately 1 percentage point at \(\ell=10\).
Among the count features in Figure~\ref{fig:selection-rate-portability}~(a), drift was relatively small for \texttt{user.followers\_count} and \texttt{user.statuses\_count}, and varied even less across lags for their corresponding rates, \texttt{followers\_rate} and \texttt{tweet\_rate}.

Even for features representing the same underlying user activity, account-age-normalised rates tended to preserve the selection rate under a source-quarter threshold better than count features at longer quarter lags.
For example, drift for \texttt{user.friends\_count} increased clearly with lag but remained relatively small for \texttt{friends\_rate}.
Drift was also smaller for \texttt{favourites\_rate} than for \texttt{user.favourites\_count}, although it remained relatively large compared with the other normalised rates.

Non-monotonic lag patterns were observed for \texttt{user.listed\_count}, \texttt{listed\_rate}, and \texttt{followers\_per\_tweet}.
In particular, drift for \texttt{followers\_per\_tweet} was larger at intermediate lags and then declined at \(\ell\geq8\).
The pairs contributing to \(\ell\geq8\) include some with similar source and target selection rates.
As in the preceding sections, the composition of quarter pairs at each lag may be associated with these lag-specific patterns.

\begin{figure}[htb]
\centering
\includegraphics[width=0.85\columnwidth]{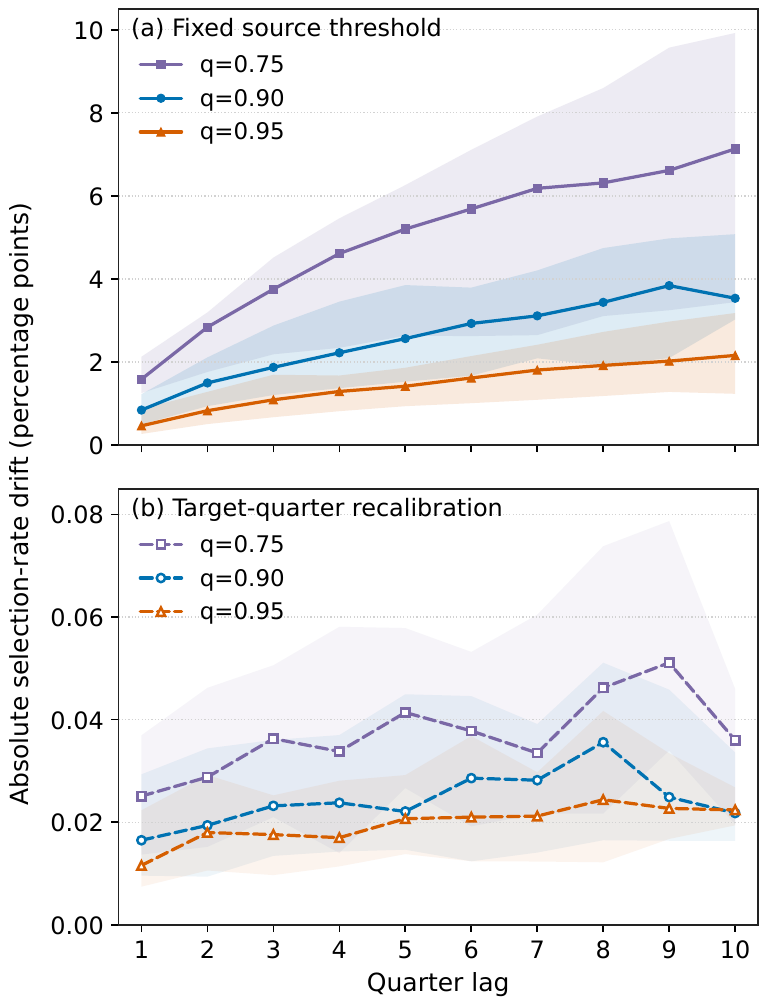}
\caption{Absolute selection-rate drift in Equations~\eqref{eq:matched-fixed-drift} and \eqref{eq:matched-recal-drift} among pairwise-common users \(C_{s,t}\).
Panel (a) shows the result when the source-quarter threshold is held fixed and applied in the target quarter, and panel (b) shows the result after target-quarter recalibration.
For \(q=0.75,0.90,0.95\), points and lines at each quarter lag show the median across feature--quarter-pair cells, and shaded regions show the interquartile range.
Panel (b) uses a different vertical-axis range from panel (a) to show the small drift after recalibration.
Recalibrating the threshold in the target quarter keeps selection-rate drift within a very small range.}
\label{fig:fixed-vs-recal-selection-rate}
\end{figure}

Figure~\ref{fig:fixed-vs-recal-selection-rate} compares absolute selection-rate drift among pairwise-common users \(C_{s,t}\) under a fixed source-quarter threshold and after target-quarter recalibration, as defined in Equations~\eqref{eq:matched-fixed-drift} and \eqref{eq:matched-recal-drift}.
When the source-quarter threshold was held fixed in Figure~\ref{fig:fixed-vs-recal-selection-rate}~(a), selection-rate drift was evident at all three quantile levels, \(q=0.75,0.90,0.95\), and tended to be larger at longer quarter lags.
By contrast, after target-quarter recalibration in Figure~\ref{fig:fixed-vs-recal-selection-rate}~(b), the lag-specific median was approximately 0.05 percentage points or less across all three quantile levels and all quarter lags, and the upper end of the interquartile range was approximately 0.08 percentage points or less.
As expected from the recalibration procedure, the selection-rate differences were close to zero, confirming that the source- and target-quarter selection rates were nearly matched.
The next section assesses whether the selected users themselves were preserved between the source and target quarters under this rate-aligned condition.

\subsection{Source-selected turnover}
\label{sec:membership}

\begin{figure*}[htb]
\centering
\includegraphics[width=\textwidth]{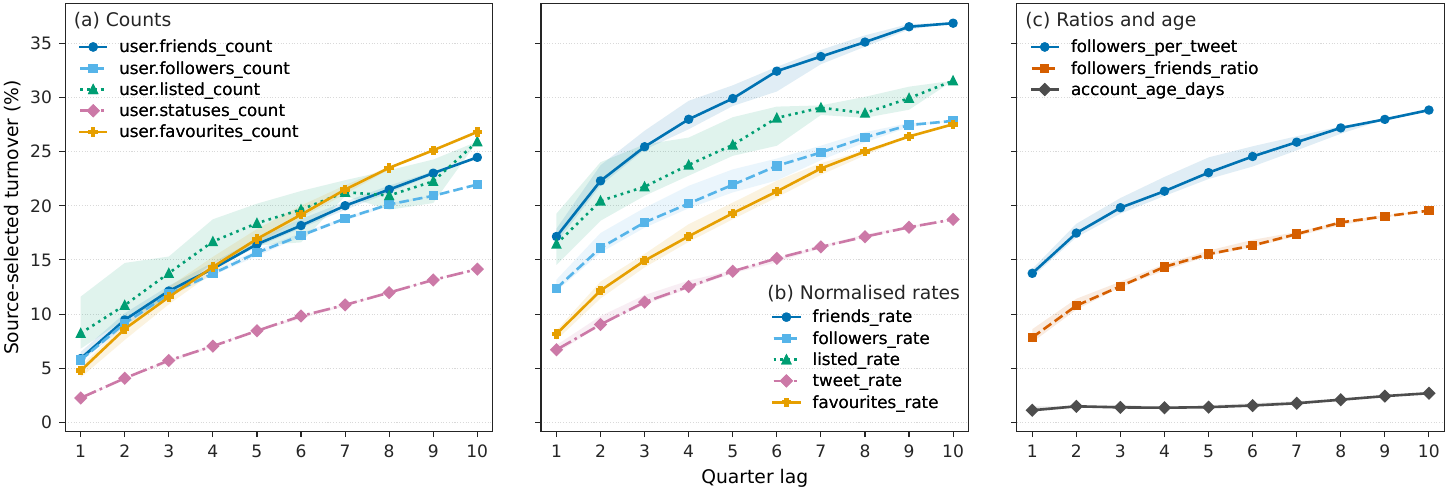}
\caption{Source-selected turnover at \(q=0.90\), expressed as the percentage of users selected in the source quarter who were not selected after target-quarter recalibration, as defined in Equation~\eqref{eq:source-selected-turnover}.
Panel (a) shows the five count features in Table~\ref{tab:user-features}, panel (b) shows the five normalised rates, and panel (c) shows ratios and age.
For each quarter lag \(\ell\), points and lines show the median of Equation~\eqref{eq:source-selected-turnover} obtained for each quarter pair over \((s,t)\in\mathcal{P}_{\ell}\), and shaded regions show the interquartile range across quarter pairs.
Source-selected turnover tends to increase towards longer quarter lags for every feature.
Turnover for \texttt{account\_age\_days} in panel (c) is small across all quarter lags, whereas particularly large turnover is observed for \texttt{friends\_rate} in panel (b), reaching approximately 17.16\% even at \(\ell=1\).}
\label{fig:source-selected-turnover}
\end{figure*}

Figure~\ref{fig:source-selected-turnover} presents source-selected turnover in Equation~\eqref{eq:source-selected-turnover} at \(q=0.90\), expressed as a percentage.
Restricting the analysis to pairwise-common users \(C_{s,t}\), we evaluated the proportion of users selected in the source quarter who were not selected after target-quarter recalibration.
As in Figures~\ref{fig:marginal-l1}--\ref{fig:selection-rate-portability}, panels (a)--(c) divide the 13 features into counts, rates, and ratios and age.

For every feature, source-selected turnover was larger at longer quarter lags.
Even at \(\ell=1\), corresponding to adjacent quarters, turnover for the 12 features other than \texttt{account\_age\_days} ranged from approximately 2.27\% for \texttt{user.statuses\_count} in Figure~\ref{fig:source-selected-turnover}~(a) to approximately 17.16\% for \texttt{friends\_rate}.
At \(\ell=10\), the corresponding values were approximately 14.2\% for \texttt{user.statuses\_count} and approximately 36.8\% for \texttt{friends\_rate}.
Thus, even after the selection rates nearly matched, selected-user membership showed clear turnover at short quarter lags, with larger turnover at longer lags.

The friends-related features showed different patterns of selection-rate drift in Figure~\ref{fig:selection-rate-portability} and source-selected turnover in Figure~\ref{fig:source-selected-turnover}.
At longer quarter lags in Figure~\ref{fig:selection-rate-portability}, fixed-threshold selection-rate drift was smaller for \texttt{friends\_rate} than for \texttt{user.friends\_count}.
By contrast, at \(\ell=10\) in Figure~\ref{fig:source-selected-turnover}, turnover was approximately 24.46\% for \texttt{user.friends\_count} and approximately 36.84\% for \texttt{friends\_rate}.
The friends-related features therefore differed in temporal portability between the fixed-threshold selection rate and selected-user membership after recalibration.

Source-selected turnover for \texttt{account\_age\_days} in Figure~\ref{fig:source-selected-turnover}~(c) was notably smaller than for the other features.
Although \texttt{account\_age\_days} increases as the observation period advances, the relative ordering of the same users was largely preserved across quarters, as shown in Figure~\ref{fig:cross-time-rank}.
Fixed-source-threshold selection-rate drift in Figure~\ref{fig:selection-rate-portability} reached approximately 6.27 percentage points at \(\ell=10\), whereas source-selected turnover after target-quarter recalibration was approximately 1.13\% at \(\ell=1\) and only approximately 2.70\% at \(\ell=10\).
Thus, for \texttt{account\_age\_days}, the feature distribution differed across quarters and fixed-source-threshold selection-rate drift was large, whereas the relative ordering of the same users was largely preserved and most source-selected users remained selected after recalibration.

Fixed-source-threshold selection-rate drift was similar for \texttt{account\_age\_days} and \texttt{user.favourites\_count} in Figure~\ref{fig:selection-rate-portability}.
By contrast, at \(\ell=10\) in Figure~\ref{fig:source-selected-turnover}, turnover was approximately 2.70\% for \texttt{account\_age\_days} and approximately 26.81\% for \texttt{user.favourites\_count}, indicating a large difference in how strongly turnover increased with lag.
Thus, even features with similar fixed-source-threshold selection-rate drift may differ substantially in the proportion of source-selected users preserved after target-quarter recalibration.

\section{Discussion}
\label{sec:discussion}

\subsection{Temporal portability of user features and selection outcomes}
\label{sec:discussion-portability}

Our results show that temporal portability depends on the property that an analysis needs to preserve.
Even for the same feature, the assessment may differ depending on whether an analysis aims to preserve the distribution of feature values, relative ordering of users, the proportion of users selected by a fixed threshold, or selected-user membership.
When metadata or a selection rule is reused in a later period, it is therefore important to specify which analytical output the subsequent analysis needs to preserve rather than simply asking whether the feature is stable over time.

This distinction was evident for several features.
For \texttt{account\_age\_days} in Sections~\ref{sec:results-marginal-distributions} and \ref{sec:results-rank-correlations}, for example, the feature distribution differed clearly across quarters, whereas the relative ordering of the same users was largely preserved.
Fixed-source-threshold selection-rate drift was also large in Section~\ref{sec:matched-cohort}, whereas source-selected turnover after target-quarter recalibration was small in Section~\ref{sec:membership}.
Thus, for \texttt{account\_age\_days}, both the feature distribution and the selection rate under a fixed source threshold differed substantially across quarters, whereas relative ordering and selected-user membership after recalibration were relatively well preserved.

Similar discrepancies occurred across features.
For example, the relative magnitudes of \texttt{friends\_rate} and \texttt{user.friends\_count} differed between fixed-threshold selection-rate drift and source-selected turnover after recalibration.
In addition, \texttt{account\_age\_days} and \texttt{user.favourites\_count} showed similar fixed-threshold selection-rate drift but differed substantially in turnover after recalibration.
A feature that is relatively well preserved with respect to one property is not necessarily preserved to the same extent with respect to another.

The distinction was also clear when a user-selection rule was reused.
Recalibration made the selection rates nearly match in Section~\ref{sec:matched-cohort}, but source-selected turnover remained in Section~\ref{sec:membership}.
Temporal portability can therefore differ between selecting the same proportion of users and selecting the same users.

\subsection{Implications for cross-time reuse}
\label{sec:discussion-implications}

The differences in temporal portability between selection rates and selected-user membership also have implications for analyses that use metadata-based user selection across periods.
If the objective is to analyse the same proportion of users in each period, recalibration using target-period data is appropriate.
If the objective is to examine the subsequent status of users selected in the source period, those users need to be fixed and followed.
If the objective is to compare the top users in each period, the selected set needs to be redefined for each quarter, and the resulting membership turnover should be treated as part of the analytical result.

For analyses that use metadata-based selection in multiple periods, it is advisable to record the feature definition, calibration period, calibration population, numeric threshold, selection fraction, and tie handling.
When selected populations are compared across periods, reporting the selection rate together with retention or turnover distinguishes changes in the proportion of users selected from changes in selected-user membership.

When the content, network, or behaviour of selected users is compared across periods, membership turnover changes the composition of the population analysed.
Such analyses should therefore examine selected-user membership as well as the selection rate.

\subsection{Scope and limitations}
\label{sec:discussion-limitations}

Digital trace data require a distinction between the observed population and the population about which inferences are made \cite{Sen2021TotalError}.
As described in Section~\ref{sec:data-collection}, we studied users observed through Japanese-language tweets in Twitter's 1\% sample stream from 2020-Q1 to 2022-Q3.
These users do not constitute Twitter's entire user population \cite{Morstatter2013SampleGoodEnough}.
Our results therefore describe temporal portability within this observation period and population.
Analyses using data from other periods or social media platforms are needed to establish the extent to which the patterns observed here apply to other observation periods, languages, or platforms.

The observation period included major social changes, including the COVID-19 pandemic.
Twitter's developer platform also changed; for example, Twitter API v2 became the primary API in November 2021~\cite{x_api_v2_primary}.
These period-specific conditions may have been associated with behaviour on Twitter and with the user population observed in each period.
Moreover, the quarter-lag summaries defined in Section~\ref{sec:user-sets} aggregate calendar-quarter pairs with the same lag.
Section~\ref{sec:results-marginal-distributions} and Appendix~\ref{app:listed-features} provide examples in which pairs containing a particular period affected a lag-specific pattern.
We did not separate the effect of the temporal interval between the source and target quarters from effects specific to individual calendar periods.
We also did not identify the effects of particular social events or platform changes on the observed temporal variation.
Comparisons of other periods or analyses restricted to particular contexts of use could examine more closely the conditions under which the observed changes arise.

Because users were observed through sampled tweets, the observed user set \(U_\tau\) in each quarter, defined in Section~\ref{sec:user-sets}, is affected by user activity, language use, and the sampling process.
Analyses using the pairwise-common user set \(C_{s,t}\), by contrast, restrict the sample to user identifiers observed in both the source and target quarters.
Our objective was to assess temporal portability; we did not quantitatively decompose the differences observed across quarters into changes in the composition of observed users and feature changes among users observed in both quarters.
Distinguishing these contributions remains a subject for future work on the factors underlying the observed temporal variation.
Moreover, an analysis using $C_{s,t}$ matches two periods, the source and target quarters; it does not continuously observe the same users in every intervening quarter or trace how they change between the two.
Future user-level longitudinal analyses could examine in greater detail how the features of users observed across periods change over time.

Finally, we examined the 13 user features defined in Section~\ref{sec:user-features} and the corresponding threshold-based user-selection rules.
The effect of the temporal variation observed here on analytical results may depend on the features, model, outcome, and task.
We leave the assessment of how the temporal variation observed here affects downstream tasks such as classification and prediction, as well as other practical settings, to future work.

\section{Conclusion}
\label{sec:conclusion}

We assessed the temporal portability of 13 user features of Twitter and the user-selection rules based on them in terms of feature distributions, same-user relative ranks, selection rates, and selected-user membership.
The analysis covered users observed through Japanese-language tweets in Twitter's 1\% sample stream from 2020-Q1 to 2022-Q3.
Feature distributions differed across quarters, and the preservation of same-user relative ranks differed by feature.
Applying a threshold defined in a source quarter in a later quarter produced selection-rate drift.
Target-quarter recalibration made the selection rates nearly match, but non-negligible changes in selected-user membership remained.
Thus, even for the same user feature, the assessment of temporal portability may depend on what an analysis aims to preserve in a later period.
When numeric user metadata and the selection rules based on them are reused across periods, temporal portability should therefore be assessed in terms of what the analysis needs to preserve.

\begin{acknowledgments}
This work was supported by the Japan Science and Technology Agency (JST) ERATO, Grant Number JPMJER2502.
M. Yasuda was supported by a Grant-in-Aid for Scientific Research from the Japan Society for the Promotion of Science (JSPS KAKENHI), Grant Number JP26K14989.
\end{acknowledgments}

\appendix
\setcounter{table}{0}
\renewcommand{\thetable}{A\arabic{table}}
\renewcommand{\theHtable}{A\arabic{table}}
\setcounter{figure}{0}
\renewcommand{\thefigure}{A\arabic{figure}}
\renewcommand{\theHfigure}{A\arabic{figure}}

\section{Data and User-Set Sizes}
\label{app:data-user-set-sizes}

Table~\ref{tab:quarterly-data} reports the number of sampled tweets and observed users in each quarter.
Table~\ref{tab:common-user-counts} reports the size of the pairwise-common user set in Equation~\eqref{eq:common-user-set} for each of the 55 source--target quarter pairs used in Section~\ref{sec:results}.
We obtained the counts in the second table by matching \texttt{user.id} from the user objects.

\begin{table*}[tb]
\caption{Number of sampled tweets and observed users by quarter.}
\label{tab:quarterly-data}
\centering
\small
\setlength{\tabcolsep}{8pt}
\begin{tabular}{@{\hspace{8pt}}llrrrr@{\hspace{8pt}}}
\toprule
Year &  & Q1 (Jan--Mar) & Q2 (Apr--Jun) & Q3 (Jul--Sep) & Q4 (Oct--Dec) \\
\midrule
2020 & Sampled tweets & 63,895,873 & 72,250,788 & 69,171,823 & 60,872,305 \\
	& Observed users & 10,130,248 & 11,010,974 & 10,782,239 & 10,075,924 \\
\midrule
2021 & Sampled tweets & 61,805,755 & 61,494,496 & 65,905,333 & 62,916,002 \\
	& Observed users & 10,324,902 & 10,272,935 & 10,681,161 & 10,568,261 \\
\midrule
2022 & Sampled tweets & 63,090,486 & 62,938,737 & 63,989,526 & \\
	& Observed users & 10,750,576 & 10,739,716 & 11,032,179 & \\
\bottomrule
\end{tabular}
\end{table*}

\begin{table*}[tb]
\caption{Cardinality of the pairwise-common user sets \(C_{s,t}=U_s\cap U_t\) in Equation~\eqref{eq:common-user-set}.
Rows give the source quarter \(s\), columns give the target quarter \(t\), and each upper-triangular cell gives \(\lvert C_{s,t}\rvert\) for \(s<t\).}
\label{tab:common-user-counts}
\centering
\scriptsize
\setlength{\tabcolsep}{2.2pt}
\resizebox{\textwidth}{!}{%
\begin{tabular}{@{}lrrrrrrrrrrr@{}}
\toprule
source \(s\) / target \(t\) & 2020-Q1 & 2020-Q2 & 2020-Q3 & 2020-Q4 & 2021-Q1 & 2021-Q2 & 2021-Q3 & 2021-Q4 & 2022-Q1 & 2022-Q2 & 2022-Q3 \\
\midrule
2020-Q1 & -- & 6,115,798 & 5,341,655 & 4,721,342 & 4,391,674 & 4,070,974 & 3,937,476 & 3,701,998 & 3,550,492 & 3,382,023 & 3,291,714 \\
2020-Q2 & -- & -- & 6,411,722 & 5,425,673 & 4,975,967 & 4,567,800 & 4,377,551 & 4,075,742 & 3,895,206 & 3,693,047 & 3,587,732 \\
2020-Q3 & -- & -- & -- & 6,225,750 & 5,507,829 & 4,984,368 & 4,742,460 & 4,382,253 & 4,167,345 & 3,933,679 & 3,808,274 \\
2020-Q4 & -- & -- & -- & -- & 6,035,330 & 5,300,395 & 4,970,161 & 4,578,380 & 4,320,944 & 4,060,657 & 3,897,261 \\
2021-Q1 & -- & -- & -- & -- & -- & 6,093,410 & 5,500,843 & 4,981,767 & 4,685,263 & 4,359,140 & 4,147,349 \\
2021-Q2 & -- & -- & -- & -- & -- & -- & 6,212,706 & 5,451,353 & 5,049,006 & 4,684,882 & 4,414,504 \\
2021-Q3 & -- & -- & -- & -- & -- & -- & -- & 6,305,339 & 5,667,569 & 5,177,427 & 4,853,567 \\
2021-Q4 & -- & -- & -- & -- & -- & -- & -- & -- & 6,271,108 & 5,577,560 & 5,162,952 \\
2022-Q1 & -- & -- & -- & -- & -- & -- & -- & -- & -- & 6,334,410 & 5,680,910 \\
2022-Q2 & -- & -- & -- & -- & -- & -- & -- & -- & -- & -- & 6,368,784 \\
2022-Q3 & -- & -- & -- & -- & -- & -- & -- & -- & -- & -- & -- \\
\bottomrule
\end{tabular}%
}
\end{table*}

\section{Quarter-Pair Patterns in Differences of Marginal Distributions}
\label{app:listed-features}

\begin{figure*}[tb]
\centering
\includegraphics[width=0.95\textwidth]{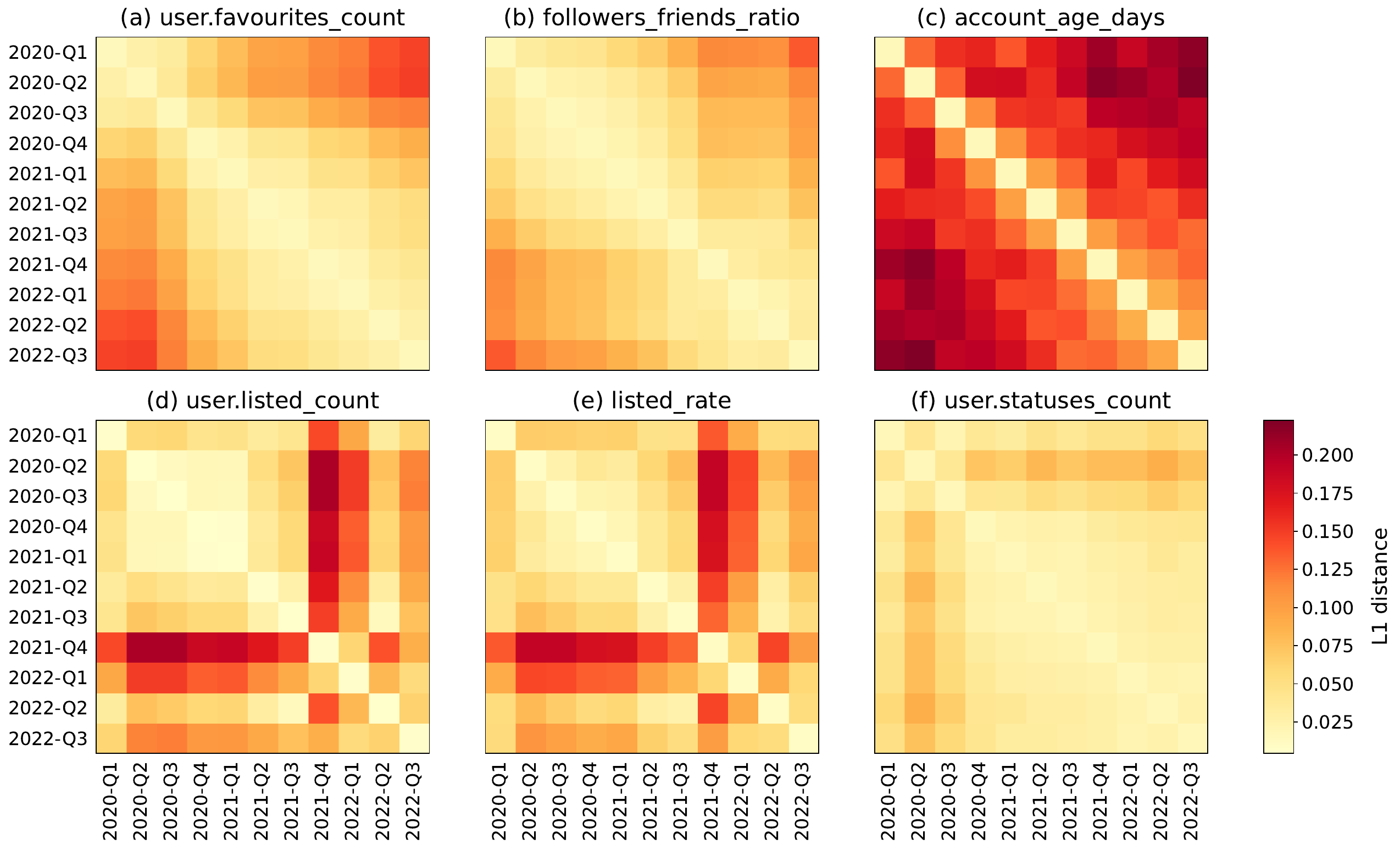}
\caption{Heatmaps of the marginal-distribution $L_1$ distance for each quarter pair for six representative features.
Rows and columns show the quarters from 2020-Q1 to 2022-Q3.
The colour of each off-diagonal cell represents the between-quarter distance defined in Equation~\eqref{eq:between-quarter-distance} for the corresponding pair of quarters. Each diagonal cell represents the within-quarter distance defined in Equation~\eqref{eq:within-quarter-distance}.
Panels (a)~\texttt{user.favourites\_count}, (b)~\texttt{followers\_friends\_ratio}, and (c)~\texttt{account\_age\_days} tend to show larger distances between quarters that are further apart in time.
In panels (d)~\texttt{user.listed\_count} and (e)~\texttt{listed\_rate}, locally large distances occur for most quarter pairs containing 2021-Q4 or 2022-Q1.
In panel (f)~\texttt{user.statuses\_count}, locally large distances occur for most quarter pairs containing 2020-Q2.}
\label{fig:period-pair-heatmap-selected6}
\end{figure*}

Figure~\ref{fig:period-pair-heatmap-selected6} presents the $L_1$ distance for each quarter pair for the two listed-related features and \texttt{user.statuses\_count}, together with three comparison features.
Each cell shows the mean $L_1$ distance for the settings in Section~\ref{sec:measure-l1}.

As described in Section~\ref{sec:results-marginal-distributions}, the between-quarter distance tends to be larger for more widely separated quarters for \texttt{user.favourites\_count} in Figure~\ref{fig:period-pair-heatmap-selected6}(a) and \texttt{followers\_friends\_ratio} in Figure~\ref{fig:period-pair-heatmap-selected6}(b).
For \texttt{account\_age\_days} in Figure~\ref{fig:period-pair-heatmap-selected6}(c), the distance is likewise markedly larger for more widely separated quarters, consistent with the definition under which values tend to increase as the observation period advances.
By contrast, although \texttt{user.listed\_count}, \texttt{listed\_rate}, and \texttt{user.statuses\_count} in Figures~\ref{fig:period-pair-heatmap-selected6}(d), (e), and (f) show a slight tendency towards larger between-quarter distances for more widely separated quarters, their patterns differ from those of the three features in panels (a)--(c).
In panels (d) and (e), locally large between-quarter distances occur for most pairs containing 2021-Q4 or 2022-Q1; in panel (f), they occur for most pairs containing 2020-Q2.

The composition of quarter pairs at each lag may also be associated with the lag-specific patterns of the listed-related features in Figure~\ref{fig:marginal-l1}.
The numbers of quarter pairs at $\ell = 5, 6, 7, 8$ are $|\mathcal{P}_{5}| = 6$, $|\mathcal{P}_{6}| = 5$, $|\mathcal{P}_{7}| = 4$, and $|\mathcal{P}_{8}| = 3$, respectively.
Of these, 2, 2, 2, and 1 pairs contain 2021-Q4 or 2022-Q1 at $\ell = 5, 6, 7, 8$, respectively.
Moreover, at quarter lags $\ell = 5, 6, 7$, 2021-Q4 or 2022-Q1 is paired with an earlier quarter, including pairs with large between-quarter distances in Figures~\ref{fig:period-pair-heatmap-selected6}(d) and (e).
By contrast, no pair at $\ell \geq 9$ contains either 2021-Q4 or 2022-Q1.
For \texttt{user.statuses\_count}, two pairs at $\ell = 1$ and one pair at each $2 \leq \ell \leq 9$ contain 2020-Q2, whereas no such pair occurs at $\ell = 10$.
The $L_1$-distance patterns for the listed-related features and \texttt{user.statuses\_count} in Figure~\ref{fig:marginal-l1} may therefore partly reflect the quarter-pair composition at each $\ell$.

\bibliographystyle{apsrev4-2}
\bibliography{references_updated_20260823}
\end{document}